# Deconstructing Anomalies in Academic Promotion & Tenure Decisions Using Spectral Graph Theory

Sanjoy Das,
Electrical & Computer Engineering,
Kansas State University,
Manhattan, KS 66506, USA.
Email: `sdas@ksu.edu`

**Abstract**: Merit based promotion & tenure decision have always been controversial. This paper suggests an agent based model of the decision making processs using spectral graph theory, where the voting agents are the vertices of the graph, and edge weights are determined based on the extent of collaborative research between the agents, as well as their estimated levels of social interactions. The model assumes that agents with lower research productivities tend to interact more often with one another. Using the graph theoretic spectrum, the paper proposes a 2-dimenional influence diagram that maps the voting agents into points on a 2-dimensional grid, where agents that are likely to influence each other more are closely spaced. Next, a dynamic model is addressed, where votes are determined based on very small randomly assigned initial values, and the mutual interaction during the decision making process. The model is able to accurately reproduce a known promotion decision making from a department of a research oriented university which involved a sizeable number of voting agents with low research outputs.

## 1. Introduction

Academic promotion and tenure (P&T) criteria can be broadly divided into two categories, merit based and seniority based (Phelan, 2004). American universities have invariably adopted merit based P&T criteria. A well devised meit based P&T policy can have a very positive impact on the department, leading to higher academic productivity (Faria & Monteiro, 2008).

Unfortunately, and in spite of P&T guidelines that individual departments and institutions formulate, such decisions are beset with controversy. According to one survey, 37% of tenure track faculty feel that their department's tenure criteria were unclear to them (American Council on Education, 2000). Studies report that personal connections between a candidate and individual members of the decision making body, significantly impacts the candidate's success rate (Gersick, Dutton & Bartunek, 2000; Zinovyeva & Bagues, 2015). Elsewhere, a theoretical study analyzes the various factors influencing uncommitted committee members and suggests that such voters significantly affect the final decision (Ayres, Rowat & Zakariya, 2007). In spite of its significance, there is a dearth of literature on theoretical models of the P&T decision process. To the best of author's knowledge, there is only one such study that models the P&T decision making process (Ayres, Rowat & Zakariya, 2007). This investigation, which is based on game theory, is largely confined to only two member committees.

A novel approach to model a departmental P&T committee using spectral graph theory (Spielman, 2007) is proposed in this research. The P&T committee and the candidate (referred to as *agents* for the remainder of this paper) are represented as the vertices of a weighted, undirected graph. There are edges connecting every pair of vertices; their weights reflecting the levels of academic as well as social interaction between the corresponding agents. The approach assumes that the degree to which agents influence each other is determined by their interactions, both research collaborations as well as other academic and social bonds.

Using spectral embedding, this paper describes how the graph's vertices can be mapped into points in a two-dimensional plane where distances between vertices depends inversely on the degrees of influence that the associated agents have upon one another. These two-dimensional representations are referred to in this paper as *influence diagrams*. The decision making process on the P&T committee is modeled next. It is assumed that the agents in the committee contain both decided and undecided voters. The votes cast by the latter are based not only on the candidate's merit but also how the agents influence each other during the decision making process. The committee agents own academic productivity is taken into account, with less productive faculty assuming more capricious positions.

Academic inbreeding, i.e. the presence of faculty with degrees from the same department, is known to have a deleterious effect on the department (Horta, Veloso & Grediaga, 2010; Inanc & Tuncer, 2011). As academically inbred faculty generally show lower research productivity levels, it has been suggested that the practice be curtailed (Horta, 2013). Academic inbreeding in briefly addressed in this research.

The results described here are based on the voting outcome of an actual department, which is kept confidential. From the actual experience of the candidate, the votes cast by the agents in that P&T committee have been estimated to a reasonably high degree of confidence. Official correspondence from competent authorities outside the department establish that the candidate merited a positive outcome, based on his/her research accomplishments. In the model described here, a high degree of emphasis is based on the candidates' research publications (Faria *et al*., 2013); consequently, productivity levels of not only the candidate agent, but also those of the P&T committee is based on the numbers of their scholarly publications. The funding levels of individual agents is also taken into account. The edge weights are determined in a similar manner, based on account research paper co-authorship and joint funding. The productivy metric is consistent with faculty seniority, with 'academic deadwood' being the least productive (Chen & Zoega, 2010; Nikolioudakis *et al*., 2015) in that particular department.

It has been shown that reasonable values of the parameters can recreate the outcome of the P&T decision with remarkable precision, thereby validating the model. The robustness of the parameters are also established here. Consistent with recent findings that inbred faculty tend to favor each other (Godechot & Louvet, 2008), small increments in the weights linking inbred agents allowed the model to reproduce the outcome to full accuracy. As one study reports the role of department leadership in productivity, although not specifically outcomes of such decision making (Bland *et al*., 2005; Parker *et al*., 2003), the role of that department's head in ensuring a fairer voting outcome has been briefly considered.

## 2. Approach

**Graph Parameters**

Academic scholarship can be determined based on a variety of factors, with research publications being the most common criterion. Other criteria include the total extramural support, the number of graduate students supervised, journal editorship, etc. Let $CRIT_k(i)$ denote the output level (e.g. publication count) of agent $i$ for criterion $k$. The weighted productivity level of agent $i$ is given by,

$$p_i = \sum_k \beta_k \frac{CRIT_k(i)}{\sum_j CRIT_k(j)}. \tag{1}$$

In the above expression, $\beta_k$ are parameters that determine the weights placed on each criterion $k$ in evaluating the overall productivity. The summation in the denominator is carried out over all agents in the department, and is used to normalize the measure, so that it never exceeds unity, and to ensure that the relative weight of criterion $k$ is entirely determined by the value assigned to the corresponding numerical value of $\beta_k$.

With $CRIT_k(i,j)$ quantifying the level of collaboration between agents $i$ and $j$ in terms of criterion $k$, the overall level of research and other academic collaboration, $R_{ij}$, is obtained using the following expression.

$$R_{ij} = \begin{cases} \sum_k \alpha_k \frac{CRIT_k(i,j)}{\sum_j CRIT_k(j)}, & i \neq j; \\ 0, & i = j. \end{cases} \tag{2}$$

| Symbol | Variable Type | Description |
|---|---|---|
| $x, X$ | Scalar | Non-bold, italics, with optional subscripts |
| **x** | Vector | Bold, lowercase, non-italics, with optional subscripts or superscripts |
| **X** | Matrix | Bold, uppercase, non-italics |
| $\mathcal{X}$ | Set | Scripts, with optional subscripts |
| $X_{rc}$ | Matrix element | Entry in row $r$ and column $c$ of matrix **X** |
| $x_r, x^r$ | Vector element | Entry in row $r$ of vector **x** |

**Table 1.** Notation.

For simplicity, the model assumes that is the amount of time that each agent $i$ devotes to scholarly work is directly proportional to the productivity, $p_i$. The remaining time can be spent in meetings as well as non-academic collaborations. Assuming that the maximum time spent by an individual within the department is $P_{\max}$, the quantity was $S_{ij}$ below determines the level of social interactions between agents $i$ and $j$.

$$S_{ij} = \begin{cases} P_{\max} - \eta \sqrt{p_i p_j}, & i \neq j; \\ 0, & i = j. \end{cases} \tag{3}$$

The quantity $\eta < 1$ is a model parameter called the socialization constant. The other parameter, $P_{\max}$ must be high enough relative to $\eta$, so that for every pair of agents $i$ and $j$, $W_{ij} > 0$. Negative values for the weights $W_{ij}$ are not usually allowed in graphs, and $W_{ij} = 0$ would result in a graph that is not fully connected.

Let $\mathcal{J}$ be the set of academically inbred agents. To capture the effect of inbred voters, the vector **u** is defined as,

$$u_i = \begin{cases} 0, & i \in \mathcal{J}; \\ 1 & i \notin \mathcal{J}. \end{cases} \tag{4}$$

The weight matrix is given by of the underlying graph is given by,

$$\mathbf{W} = \mathbf{R} + \mathbf{S} + \gamma \mathbf{u}\mathbf{u}^T. \tag{5}$$

The quantity $\gamma$ is another model parameter called the inbreeding constant. To neglect any extra amount of influence that agents in $\mathcal{J}$ may exert upon one another, the value of $\gamma$ may be optionally set to zero, although a small positive value is suggested. The Laplacian **L** of the graph is obtained as,

$$L_{ij} = \begin{cases} -W_{ij}, & i \neq j; \\ \sum_k W_{ik}, & i = j. \end{cases} \tag{6}$$

With $N$ being the total number of agents, **W** and **L** are $N \times N$ symmetric matrices.

**Influence Diagram**

The Laplacian **L** has $N$ eigenvectors, $\mathbf{v}_n$ ($n = 1, 2, \ldots, N$). Each eigenvector has an associated eigenvalue $\lambda_n$. The eigenvalues and eigenvectors are indexed so that $\lambda_1 \leq \lambda_2 \leq \cdots \leq \lambda_N$. The smallest eigenvalue $\lambda_1$ of any graph is always zero. The two-dimensional influence diagram is a plot of $\lambda_2^{-1}\mathbf{v}_2$ (x-axis) vs. $\lambda_3^{-1}\mathbf{v}_3$ (y-axis). The coordinates of agent $i$ in this plot are the $i^{\text{th}}$ entries of the vectors $\lambda_2^{-1}\mathbf{v}_2$ and $\lambda_3^{-1}\mathbf{v}_3$. Since all the weights of the graph are strictly positive, none of the other eigenvalues are zero, in particular $\lambda_2, \lambda_3 > 0$.

The primary purpose of the influence diagram is for visualization. Agents that collaborate more often appear closer together in the diagram. Visualization can be further enhanced by quantifying the amount of influence any agent $i$ with coordinates $\mathbf{z}_i$ exerts at any other point $\mathbf{z}$ in the plot. For example, if $x_i$ is the voting decision of each agent $i$, the net influence $x(\mathbf{z})$ at $\mathbf{z}$ of all agents can be formulated using the following expression.

$$x(\mathbf{z}) = \sum_i x_i e^{\frac{1}{\sigma}\|\mathbf{z} - \mathbf{z}_i\|^2}. \tag{7}$$

In the above expression, $\|\mathbf{z} - \mathbf{z}_i\|$ is the Euclidean distance between the two dimensional points $\mathbf{z}$ and $\mathbf{z}_i$, and the quantity $\sigma$ is a small parameter that can be adjusted for best visualization.

**Dynamic Voting Analysis**

In a typical P&T decision process, some of the agents in the committee enter the process having determined *a priori* there voting decision. Their participation is restricted to influence the votes of the undecided agents (Ayres, Rowat & Zakariya, 2007). Accordingly, the model distinguishes between three sets of agents, $\mathcal{V}^-$ and $\mathcal{V}^+$ are the sets of agents with prior decisions to vote against, and in favor of the candidate. The set of undecided agents, $\mathcal{V}_U$ comprises of all remaining agents. Note that the candidate agent $c$ is not included in any of these sets, $\mathcal{V}^-$, $\mathcal{V}^+$ or $\mathcal{V}_U$.

The votes cast by each undecided agent is based on a decision variable $x_i$. When $x_i$ is positive, the agent $i$ votes in favor of the candidate, and when it is negative, the agent votes against the latter. For agents in $\mathcal{V}^-$ and $\mathcal{V}^+$ the decision variable is only used to influence the undecided agents. Prior to the decision process, each agent's decision variable in $\mathcal{V}_U$ is assigned a small random value $x_i^0$. For the agents in $\mathcal{V}^-$ and $\mathcal{V}^+$, the values of $x_i^0$ are set to $-1$ and $+1$. For the undecided agents, the initial value is obtained in the following manner.

$$x_i^0 = \alpha \left( \frac{p_i}{P_{\max}} m_c + \left(1 - \frac{p_i}{P_{\max}}\right) r \right), \qquad i \in \mathcal{V}_U. \tag{8}$$

In the above expression, the quantity $r$ is a uniformly distributed random variable lying in the range $-1$ to $+1$ (i.e. $r \sim U[-1, +1]$) and $\alpha$ is a small positive constant that determines the maximum randomness. The quantity $m_c \in [-1, +1]$ quantifies the overall merit of the candidate. The factor appearing within parenthesis to the right of the above equation contains two terms. The first term is directly proportional to $p_i$ so an undecided agent with a higher productivity places greater emphasis on the candidate's own merit worthiness. The second term allows the votes cast by agents with lower productivities acquire more random initial values. It should be noted that the only role of the noise parameter $\alpha$ is to determine the range of initial voting assignments $x_i^0$ of the undecided agents in $\mathcal{V}_\text{U}$.

For a deserving candidate the quantity $m_c$ can be set to a value of at most +1; conversely for an undeserving candidate $m_c$ can be assigned a value of no less than –1. Unless the merit of the candidate can be determined with a very high degree of confidence through extraneous means, a suggested way to assign a value to $m_c$ is according to either of the following two expressions provided below.

$$m_c = \frac{1}{P_\text{max}}\left(p_c - \frac{1}{2}P_\text{max}\right), \tag{9}$$

or,

$$m_c = \frac{1}{P_\text{max}}\left(p_c - \frac{1}{2N}\sum_i p_i\right). \tag{10}$$

The final value of the decision variables is determined in accordance with the expression shown below.

$$\mathbf{x} = ((1+\varepsilon)\mathbf{I} + \mu\mathbf{L})^{-1}\mathbf{x}_0. \tag{11}$$

The quantities $\mu$ and $\varepsilon$ above are the influence and regularization constants. The influence constant sets the degree to which agents' decisions are decided through mutual interactions, relative to their random initial assignments. It can be shown that the above expression for $\mathbf{x}$ minimizes the following cost function.

$$\varphi(\mathbf{x}) = \frac{1}{2}(\|\mathbf{x}-\mathbf{x}_0\|^2 + \varepsilon(\mathbf{x}-\mathbf{1})^\text{T}(\mathbf{x}+\mathbf{1}) + \mu\mathbf{x}^\text{T}\mathbf{L}\mathbf{x}). \tag{12}$$

The first term within parenthesis appearing to the right of the above expression for the cost function acquires a minimum value of when none of the agents in the committee deviate from their initially assumed values, i.e. when $\mathbf{x} = \mathbf{x}_0$. This it minimizes the deviations of the agents' decisions from their initial values. The second term is zero when each $x_i$ is either $+1$ or $-1$. It is optional, and can be included if the voting outcomes be close to $\pm 1$. Simulations indicate that the relative outcome is not affected even with the constant $\varepsilon$ being unity, although a sufficiently small value is recommended.

The third term in the expression for $\varphi(\mathbf{x})$ involves the Laplacian $\mathbf{L}$. It can be shown that it simplifies as follows.

$$\mathbf{x}^\text{T}\mathbf{L}\mathbf{x} = \frac{1}{2}\sum_i\sum_{j\neq i}W_{ij}(x_i - x_j)^2. \tag{13}$$

The above expression shows that $\mathbf{x}^\text{T}\mathbf{L}\mathbf{x}$ is the sum of the squared difference between the decisions of every pair of agents $i$ and $j$, weighted by $W_{ij}$ which depicts the amount of influence they have on each other. Thus, each term tries to keep the voting outcomes of agents that exert more influence on each other, to remain closer. The quantity $\mu$ is termed the mutual influence constant.

## 3. Case Study

**Committee Makeup**

As the scenario considered here pertained to the promotion decision of a candidate, multiple years were involved. Hence, the voting decisions of the committee members could be estimated to a very high degree of confidence. There were 11 agents, labeled 1 – 11, including the candidate ($c = 1$). The department head ($h = 4$) was absent from the P&T committee, which consisted of the remaining 9 agents, so that $\mathcal{V} = \{2,3,5,6,7,8,9,10,11\}$. The agents in the P&T committee who made their decisions prior to the committee consisted of two supporting agents, who have a strong history of collaborative research with the candidate (7,8), and two malevolent agents (6,11) who cast negative votes. The set of academically inbred agents, who had received at least one degree from the same department, was $\mathcal{I} = \{2,3,4,11\}$.

The justification that was provided by agents 6 and 11 for their negative votes is considered in this research as not suitable grounds for that department's P&T decisions because of three reasons: (*i*) the posited rationale they provided were not listed in the department's P&T guidelines, (*ii*) the candidate's research was subsequently found to meet or exceed the promotion criteria by more competent, higher level authority outside the department, and significantly, (*iii*) the previous year, another candidate (10) had been promoted without meeting the same requirements, and with both 6 and 11 voting in favor. For these reasons, the candidate has been considered as meriting a positive outcome wherever needed in the results described later. Additionally, it should be noted that the productivities, $p_6 = 0.789$ and $p_{11} = 0.727$ of agents 6 and 11 were found to be relatively low in comparison to others, and based on their productivities as well as retirement status, agents 6 and 11 may be considered academic deadwood (Chen & Zoega, 2010). Conversely, that of agents 7 and 8 were among the most productive in the department ($p_7 = 4.087$ and $p_8 = 2.769$) (see below).

Furthermore, agent 10 who had been successfully promoted the previous year, also voted against the candidate, agent 1's P&T decision, despite being clearly aware that the rationale being set forth by agents 6 and 11, during the P&T meeting were not appropriate reasons. This agent (10) is classified as a strategic agent – the equivalent of a zero-sum player within a game theoretic context. Agent 10 also subsequently received academic recognition for which the candidate, agent 1, would have also qualified. It is recommended that without strong underlying reasons, such an agent should be included within the set of undecided agents.

The final voting sets obtained is this manner were the set of malevolent and strategic agents, $\mathcal{V}^- = \{6,10,11\}$, the set of supportive agents, $\mathcal{V}^+ = \{7,8\}$, and the set of undecided agents $\mathcal{V}_U = \{2,3,5,9\}$, which includes the graduate program coordinator ($g = 9$). The simulations detailed below focus on how these four prior undecided agents arrived at their eventual decisions. Although to the best of the author's knowledge, the voting decisions of all agents as well as their underlying justifications, as described earlier, are authentic, the simulations below are generalized enough to be able to recreate any other similar P&T decision process with a high degree of fidelity. The parameters used in the model were few, and they were assigned reasonable numerical values (sometimes zero). The model was found to be very robust to changes in these values.

**Graph Parameters**

The productivity vector **p** and the collaboration matrix **R** were constructed using only data that was readily available online at the institution's website, which consisted of the number of journal publications, the total number of publications (including journal, conferences, and technical reports), and the funding obtained. A period of 4 years prior to the candidate's P&T application was considered. If $JOUR(i)$, $PUBL(i)$, and $FUND(i)$ represent the total number of journal articles published, total number of publications, and the total grant level by agent $i$ during that period, its productivity level was computed in the manner suggested in Eq. (1), as follows.

$$p_i = 0.67 \times \frac{JOUR(i)}{\sum_k JOUR(k)} + 0.33 \times \frac{PUBL(i)}{\sum_k PUBL(k)} + \frac{FUND(i)}{\sum_k FUND(k)}. \quad (14)$$

Note that for jointly funded projects, the quantity $FUND(i)$ was divided by the total number of PIs and co-PIs involved. In accordance with other findings (Faria *et al.* 2013), the journal publication count is assigned a higher weightage in comparison to other research publications. The numerical value of the productivity vector obtained was,

$\mathbf{p} = [4.127\ 0.229\ 0.930\ 1.788\ 3.779\ 0.789\ 4.087\ 2.769\ 0.138\ 2.637\ 0.727]$.

In order to account for administrative responsibilities of the department chair ($h = 4$) and the graduate coordinator ($g = 9$), their productivities were incremented, so that the adjusted productivities were,

$$p_i = \begin{cases} p_i + 2, & i = h; \\ p_i + 1, & i = g; \\ p_i, & \text{otherwise}. \end{cases} \quad (15)$$

These adjusted values were used in all simulations described in this research.

With $JOUR(i,j)$ and $PUBL(i,j)$ denoting the total number of published journal articles and all articles co-authored by any pair of agents $i$ and $j$, and $FUND(i,j)$ being their joint funding, the academic and research collaboration between them was quantified in the following manner (see Eqn. (2)).

$$R_{ij} = \begin{cases} 0.67 \times \frac{JOUR(i,j)}{\sum_k JOUR(k)} + 0.33 \times \frac{PUBL(i,j)}{\sum_k PUBL(k)} + \frac{FUND(i,j)}{\sum_k FUND(k)}, & i \neq j; \\ 0, & i = j. \end{cases} \quad (16)$$

The collaboration matrix $\mathbf{R}$ found using the above expression is shown below.

$$\mathbf{R} = \begin{bmatrix} 0.000 & 0.000 & 0.000 & 0.000 & 0.000 & 0.000 & 0.891 & 1.363 & 0.000 & 0.441 & 0.000 \\ 0.000 & 0.000 & 0.016 & 0.000 & 0.000 & 0.000 & 0.185 & 0.000 & 0.000 & 0.000 & 0.000 \\ 0.000 & 0.016 & 0.000 & 0.000 & 0.000 & 0.000 & 0.000 & 0.000 & 0.000 & 0.000 & 0.000 \\ 0.000 & 0.000 & 0.000 & 0.000 & 0.063 & 0.000 & 0.812 & 0.436 & 0.000 & 0.385 & 0.398 \\ 0.000 & 0.000 & 0.000 & 0.063 & 0.000 & 0.000 & 0.526 & 0.146 & 0.252 & 0.000 & 0.000 \\ 0.000 & 0.000 & 0.000 & 0.000 & 0.000 & 0.000 & 0.000 & 0.101 & 0.000 & 0.000 & 0.000 \\ 0.891 & 0.185 & 0.000 & 0.812 & 0.526 & 0.000 & 0.000 & 0.371 & 0.000 & 0.049 & 0.000 \\ 1.369 & 0.000 & 0.000 & 0.436 & 0.146 & 0.101 & 0.371 & 0.000 & 0.117 & 0.016 & 0.000 \\ 0.000 & 0.000 & 0.000 & 0.000 & 0.252 & 0.000 & 0.000 & 0.117 & 0.000 & 0.000 & 0.000 \\ 0.441 & 0.000 & 0.000 & 0.385 & 0.000 & 0.000 & 0.049 & 0.016 & 0.000 & 0.000 & 0.000 \\ 0.000 & 0.000 & 0.000 & 0.398 & 0.000 & 0.000 & 0.000 & 0.000 & 0.000 & 0.000 & 0.000 \end{bmatrix}.$$

**Influence Diagram**

Figure 1 shows the two-dimensional influence diagram obtained from the 2nd and 3rd smallest eigenvectors of the Laplacian, for the following values, $\eta = 0.20, 0.35, 0.50$ and $P_{\max} = 6, 9, 12$ and $\gamma = 0.0$. Figure 2 is the influence diagram with the $\eta = 0.25$, $P_{\max} = 7$, and $\gamma = 0.25$ that were used in all remaining simulations. The value $P_{\max} = 7$ can be interpreted as the productivity of a hypothetical agent who works seven days a week. The numerical values in figures 1 and 2 are slightly different merely to illustrate the robustness of the influence diagram. In both figures, the points are color coded for better visualization. The candidate (agent 1) is shown in yellow (bottom left). The agents in $\mathcal{V}^+$ as well as agent 9 (included in $\mathcal{V}_U$) who voted favorably appear in blue, whereas the agents that voted against are in red. The department chair who did not participate is shown in green. Agent 9, who voted in favor, and against the candidate in two separate instances, is shown in magenta.

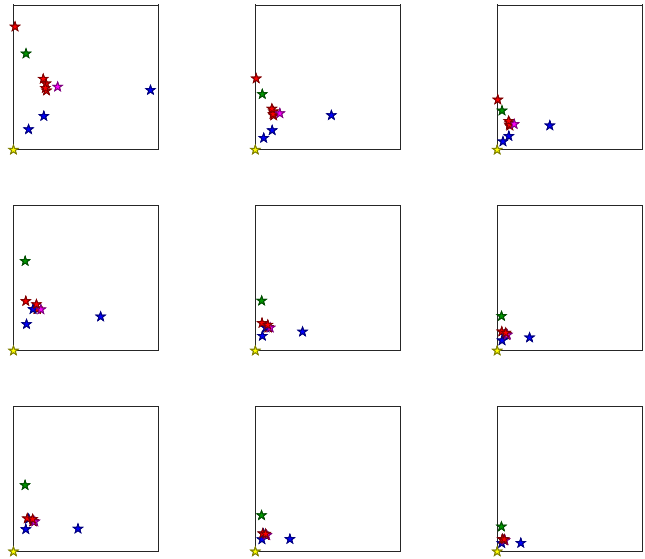

**Figure 1**. Influence diagram with $\eta = 0.2$ (top row), 0.35 (middle row) and 0.5 (bottom row) and $P_{\max} = 6$ (left column), 9 (center column) and 12 (right column). In all plots, $\gamma = 0$.

The same scale was used in all nine plots in figure 1. As $P_{\max}$ is increased, the points move closer to each other. This is because higher $P_{\max}$ results in higher values to the entries in the matrix **S**, representing social interactions. This can be interpreted as the agents having more time for socialization, thus increasing their levels of non-research related collaborative activity. The directions of movement of the points as $\eta$ is varied can be interpreted in a similar manner. It is also clearly observed that when $P_{\max}$ and $\eta$ are varied, excluding agents 4 and 9, the relative positions of the other agents remain largely unchanged. This highlights the robustness of the influence diagram with respect to $P_{\max}$ and $\eta$. The points 4 and 9, which represent the department chair and graduate coordinator change relative to those of the other agents because their productivities $p_4$ and $p_9$ are incremented by constant amounts whereas those of the other agents are allowed to change in accordance with $P_{\max}$ and $\eta$.

The proximity of the points representing the undecided agents 2 and 3 to two malevolent agents 6 and 11 clearly shows how agents 2 and 3 voted negatively. Agent 9 is at a higher distance, and consequently did not vote negatively always. Agent 5 who is placed at a significantly larger distance, was clearly not influenced by the malevolent agents, explaining its eventual decision to cast a vote in favor. The effectiveness of the influence diagram in predicting the agents' decisions is clearly demonstrated.

Although the chair (agent 4) did not participate in the P&T decision making process, its potential role in influencing the choices of the other agents can be investigated through influence diagrams. This is addressed in the contoured plots in figure 3 where the agents in $\mathcal{V}^+$ are assigned voting decisions $x_i = +1$ while those in $\mathcal{V}^-$, $x_i = -1$. For the undecided agents in $\mathcal{V}_U$, $x_i = 0$. The contours are based on Eq. (7) with $\sigma = 4 \times 10^{-4}$. The decision $x_4$ of agent 4, representing the chair is kept at $x_4 = 0$ in the left plot as the chair did not participate in the P&T committee during the decision making process. The contours show how the agents who had made decisions *a priori* affected the others with pure blue and red colors corresponding to the extreme decisions of +1 and −1. Within the square region, only 28% have values $x(\mathbf{z}) > 0$. In the middle plot, $x_4 = +1$ to show how the influences change with agent 4's presence and playing an equal role as the other decided agents. The net positive area is now increased to 42%. The plot to the right depicts what would happen if the chair agent assumes a larger role than the other agents, depicting effective leadership. Accordingly, $x_4 = +2$ and $\sigma$ is increased to $16 \times 10^{-4}$ for this agent. The area covered by $x(\mathbf{z}) > 0$ increases significantly to 94% showing that the presence of an effective chair can play a decisive role in ensuring a fairer outcome. Although not directly related to this situation, the positive role of effective leadership has been reported elsewhere (Bland *et al.*, 2005).

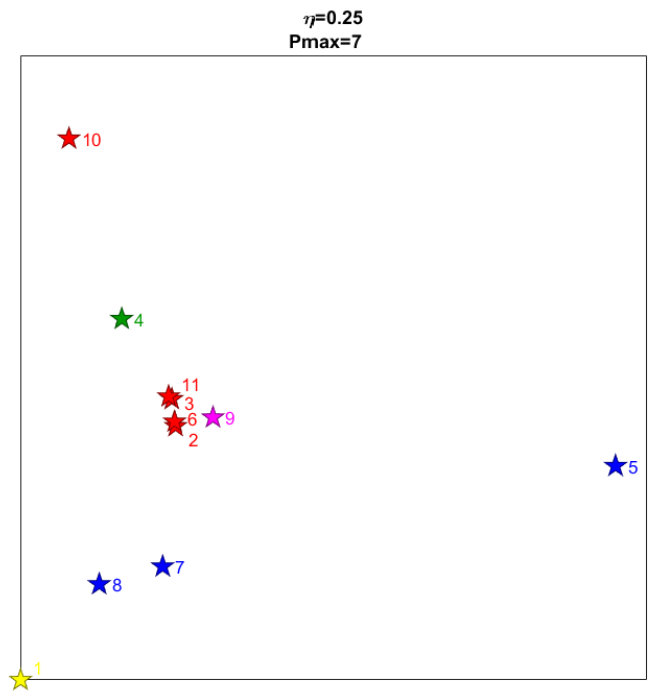

**Figure 2**. Influence diagram with $\eta = 0.25, P_{\max} = 7, \gamma = 0.25$.

**Dynamic Voting Analysis**

While the analysis so far focuses entirely on how the agents influence each other based on the Laplacian, **L**, the study discussed next, takes into account the productivity vector **p** and is based on Eq. (12). The decision variable $x_i^0$ was initialized as shown in Eq. (8). The value of $\alpha$ was low at $\alpha = 0.15$, and that of $\varepsilon$ was kept at $\varepsilon = 0$. With all other parameters at their earlier values, this study examined the role of inbreeding in influencing the undecided agents. Therefore $\gamma$ was varied between 0 and 1 in steps of 0.2. The final decision vector was obtained for each value of $\gamma$, as shown in figure 4 with $\mu = 0.5$ (left), $\mu = 1$ (middle), and $\mu = 2.0$ (right). Within each subplot, the decisions of the agents in $\mathcal{V}_U$ appear sorted in order of increasing productivities. Thus, the decision $x_2$ of agent 2 ($p_2 = 0.229$) is the leftmost and $x_5$, which is that of agent 5 ($p_5 = 3.779$) is the rightmost. Within each histogram, the vertical bars are colored according to the value of $\gamma$. The purple bars correspond to $\gamma = 0$, blue bars to $\gamma = 0.2$ and so on, until the yellow bars, which are for $\gamma = 1$.

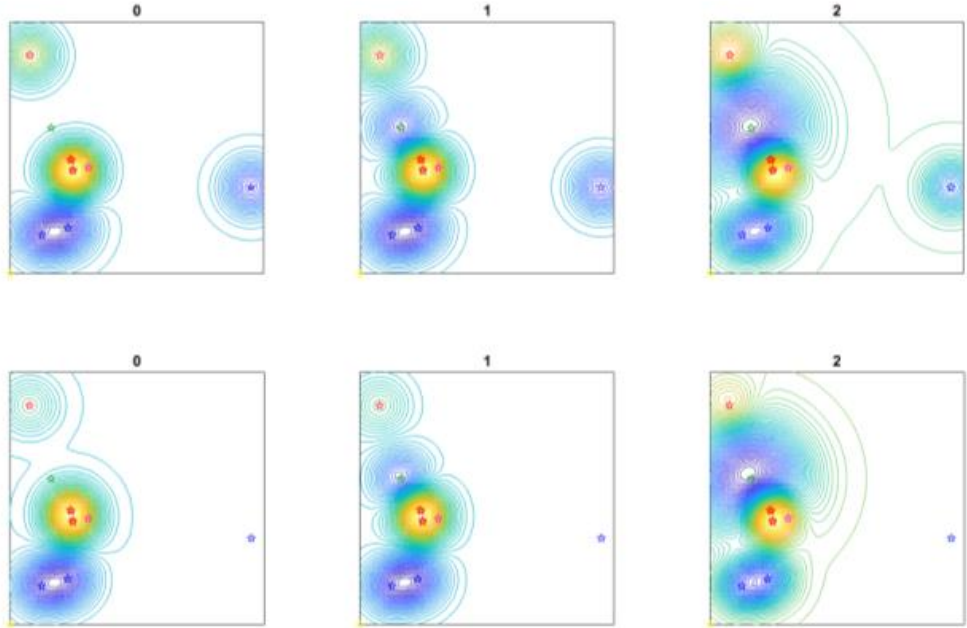

**Figure 3**. Influence diagram showing how an effective department chair may influence the outcome of the decision process with the following values of the constants: $\eta = 0.25$, $P_{\max} = 7$, $\gamma = 0.25$. The quantity $\sigma = 4 \times 10^{-4}$ for all agents except the department chair (4) which was $\sigma = 4 \times 10^{-4}$ (left and middle columns) and $\sigma = 16 \times 10^{-4}$ (right column).

The effect of increasing $\gamma$ is clearly seen. As it increases, the decisions $x_2$ and $x_3$ decrease steadily. This is because agents 2 and 3, which are in $\mathcal{J}$ are more influenced by malicious agent 11 who is also in $\mathcal{J}$. The decisions of the other agents, agent 5 and agent 9 remain largely unaffected. This decrease is seen for all three values of $\mu$. Moreover, the similarity of the three subplots indicate the robustness of the model to the parameter $\mu$.

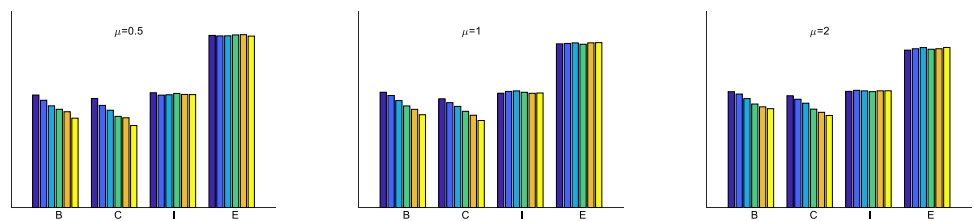

**Figure 4**. Histograms showing the average votes of the undecided candidates, sorted in order of productivity with $\mu = 0.5$ (left), $\mu = 1.0$ (middle) and $\mu = 2.0$. (right). The inbreeding parameter $\gamma$ were at values $\gamma = 0, 0.2, 0.4, 0.6, 0.8, 1.0$.

## 4. Conclusion

This research accurately models the outcome seen in a recent P&T decision process at a department. The proposed influence diagram which is based on algebraic graph theory, very effectively captures the voting patterns of the undecided agents. The dynamic analysis was able to eliminate the minor discrepancy that persisted between the model's prediction with the observed outcome through the introduction of a small inbreeding coefficient. Although not shown, preliminary results in applying the model to randomly generated data reveals that the model is able to faithfully reproduce 'common sense' results where more research oriented faculty vote more consistently and fairly, than less productive ones. The author is investigating if there exists a tipping point in the ratio of unproductive to productive faculty, that when exceeded, leads to anomalous outcomes.